\documentclass{article}
\usepackage{epsf}
\usepackage{graphicx}
\def\beq{\begin{equation}}
\def\eeq{\end{equation}}
\begin{document}
\title{\bf Calculations of time-dependent observables in non-Hermitian
quantum mechanics: The problem and a possible solution }
\author{Ido Gilary, Avner Fleischer and Nimrod Moiseyev}
\maketitle
\begin{center}
{ Department of Chemistry and Minerva Center for Nonlinear Physics
of Complex Systems, Technion -- Israel Institute of Technology,
Haifa 32000, Israel. \\}
\vspace{1.0cm}
\end{center}
\begin{abstract}

The solutions of the time independent Schr\"odinger equation for
non-Hermitian (NH) Hamiltonians have been extensively studied and
calculated in many different fields of physics by using $L^2$
methods that originally have been developed for the calculations
of bound states. The existing non-Hermitian formalism breaks down
when dealing with wavepackets(WP). An open question is how time
dependent expectation values can be calculated when the
Hamiltonian is NH ? Using the F-product formalism, which was
recently proposed, [J. Phys. Chem., {\bf 107}, 7181 (2003)] we
calculate the time dependent expectation values of different
observable quantities for a simple well known study test case
model Hamiltonian. We carry out a comparison between these results
with those obtained from conventional(i.e., Hermitian) quantum
mechanics (QM) calculations. The remarkable agreement between
these results emphasizes the fact that in the NH-QM, unlike
standard QM, there is no need to split the entire space into two
regions; i.e., the interaction region and its surrounding. Our
results open a door for new type of WP propagation calculations
within the NH-QM formalism that until now were impossible. In
particular our work is relevant to the many different fields in
physics and chemistry where complex absorbing potentials are
introduced in order to reduce the propagation calculations into a
restricted region in space where the artificial reflections from
the edge of the numerical grid/box are avoided.
\end{abstract}

\section{Introduction} \label{intro}

In the last two decades there has been an increasing interest in
non-Hermitian(NH) quantum mechanics (QM) in many different fields
of physics. In many studied cases in atomic, molecular and nuclear
physics the Hamiltonian is NH due to the use of different type of
analytical continuation transformations. One such method is
complex scaling (CS) where $\hat{H}({\bf r})\to \hat{H}({\bf
r}e^{\imath\theta})$
\cite{Junker-review}\cite{Renihardt-review}\cite{Ho-review}\cite{NM-review}.
Another common method is the introduction of complex absorbing
local energy independent potentials (CAPs), also known as optical
potentials, into the
Hamiltonian\cite{Caps}\cite{lenz-robin-review} such that $\hat{H}
\to \hat{H} + \hat{V}_{CAP}$. The complex eigenvalues and
eigenfunctions of the NH Hamiltonians, in these studies, were
calculated using square-integrable methods that were originally
developed for the calculation of bound states (see references to
reviews  mentioned above). The need to replace the scalar product
which is used in conventional QM by a generalized inner product is
well understood and has been discussed quite extensively in the
literature. See for example the different approaches for
generalized inner products in NH-QM in Ref. \cite{zeldovich} and
Ref.\cite{Letropet} and references therein. However, for the sake
of clarity in Section \ref{c-product} we provide a brief account
of the generalized inner product (so called
c-product\cite{Mol-Phys-79}) which has been used until now for
calculating time independent observables in NH-QM\cite{NM-review}.
It is quite straightforward to apply the same generalized
definition of the inner product to NH time periodic Hamiltonians (
see for example Ref.\cite{NM-review}), which describe the
interaction of atoms or molecules with oscillating electric field.
The application of the NH formalism to laser driven system has
been very useful in explaining physical phenomena. For example,
remarkable agreement has been achieved between the results
obtained from theoretical NH-QM calculations and the experimental
results of the probability of generating odd high-order harmonics
(i.e., generation of high energy photons resulting from the
absorbtion of many low energy photons)\cite{NM-FW-PRL}. However,
one important problem remains open. How to calculate the entire
high-order harmonic generation spectra and not only the
probabilities to obtain even or odd harmonics ? This question is
related to a more general question. How to calculate time
dependent expectation values in NH-QM ? In
 NH-QM the propagated right wavepacket (WP) is the solution of the time dependent
Schr\"odinger equation (TDSE) where $\hat{H}$ is a NH Hamiltonian.
\begin{equation}
\hat{H}\Psi({\bf r},t)=\imath \hbar \frac{\partial}{\partial
t}\Psi({\bf r},t) \quad. \label{NH-TDSE}
\end{equation}
Extensive WP propagation calculations were carried out for complex
potential energy surfaces by solving the NH-TDSE (see for example
Refs.\cite{kosloff,Leforestier-Wyatt,Balint-Kurti,Dieter-Mayer,NM-SS-LSC}).
However, in order to calculate time dependent expectation values
one needs to carry out WP propagation calculations from $t=0$ to
$t=-\infty$. The problem is that in NH-QM it is impossible to
propagate the initial WP from t=0 to $t=-\infty$. We briefly
review this known problem in Section \ref{asymmetry} and relate it
to the concept of time asymmetry in QM which is a subject of
numerous studies in the literature (see
Refs.\cite{Bohm,Prigogine,Sudarshan,Nicolaides} and references
therein).\\
Very recently a possible solution to the time asymmetry problem in
NH-QM  was proposed by Moiseyev and Lein\cite{NM-Lein}. The
solution requires modification of the currently used inner
product. As we will show in Section \ref{F-product} this
modification of the well known generalized inner product (briefly
reviewed in Section \ref{c-product}) which we will term
"F-product" (FP) is applicable for cases where the Hamiltonian is
either a time-independent or time dependent NH operator. Naturally
this modification of the generalized inner product reduces to the
conventional scalar product when the Hamiltonian is Hermitian.
Moiseyev and Lein \cite{NM-Lein} have shown that using the
modified generalized definition of the inner product not only
solves the time asymmetry problem in NH-QM , but also has a very
important physical implications. They have shown that the
analytical expression for the high-order harmonic generation
spectra (HHGS) which has been derived within the framework of the
FP modified NH time dependent formalism, clearly indicates why
only odd high-order harmonics were observed in experiments where
high-intensity laser pulses have been used. Moreover, the analysis
of the expressions they obtained for the HHGS indicates under what
conditions also even high-order harmonics would be observed.
Simulation calculations within the framework of the conventional
(hermitian) QM based on their finding supported their results. The
possibility of generating also the even-order harmonics as well as
the odd ones should help in generating ultra short high intensity
laser pulses. However, so far the FP generalized inner product as
first proposed in ref\cite{NM-Lein} has never been tested for
numerical NH calculations. The purpose of this work is to
calculate time-dependent observable quantities for a well known
test case study problem within the framework of the FP formalism
and compare it with the results obtained from the conventional
(i.e., Hermitian) QM. As we will show here such a comparison
clearly shows that time dependent observables should be calculated
within the framework of the FP formalism and also give some
physical interpretation of this formalism.
\section{Non-Hermitian Hamiltonians, and their consequence}
\label{resonances} Non-Hermitian Hamiltonians have been used to
describe a variety of physical phenomena. The methods by which the
Hamiltonian becomes NH are also diverse. Such methods include for
example the addition of complex absorbing potentials to the
Hamiltonian or the scaling of the coordinate by a complex factor.
The use of NH Hamiltonians results in complex eigenvalues,
\begin{equation}
E_\alpha=\varepsilon_\alpha-\imath \Gamma_\alpha/2 \quad,
\end{equation}
 where often the real part,
$\varepsilon_\alpha=\textit{Re}(E_\alpha$), is associated with the
energy while the complex part is related to the decay rate,
$\Gamma_\alpha$ of a metastable state by:
$\Gamma_\alpha=-2{\textit{Im}}(E_\alpha)$ such that the lifetime
of the metastable state is given by $\hbar/\Gamma_\alpha$. When
using NH Hamiltonians one needs to address the aspects that differ
from conventional QM.
\subsection{Brief account of the generalized inner product: the
c-product} \label{c-product} As a result of the NH nature of the
Hamiltonian, there is a need to define a generalized inner
product. The question of the definition of the inner product when
the Hamiltonian is NH is a crucial one, especially in time
dependent calculations. Let us consider  a general case where the
Hamiltonian, $\hat {H}$ can be either Hermitian or NH Hamiltonian.
Since the concept of a generalized inner product is well defined
in linear algebra for general not necessarily Hermitian
matrices\cite{wilkinson} we begin our brief review  by the
representation of the Hamiltonian by a matrix $\bf H$. The
Hamiltonian matrix elements are defined as
\begin{equation}
H_{i,j}=<\varphi_i|{\hat H}|\varphi_j> \label{HAM}
\end{equation}
where $\{\varphi_i\}$ are orthonormal square integrable basis set
functions in the Hilbert space. We assume here that we can
truncate the number of the basis functions and represent the
Hamiltonian by a finite general complex matrix. The matrix can be
as large as one wishes. The matrix $\bf H$ has right and left
eigenvectors. Let us denote the right (column) eigenvectors by
${\vec \Phi}_\alpha^R$ and the left (row) eigenvectors by $({\vec
\Phi}_{\alpha'}^L)^t$. That is,
\begin{equation}
{\bf H}{\vec \Phi}_\alpha^R=E_\alpha {\vec \Phi}_\alpha^R
\end{equation}
and
\begin{equation}
({\vec \Phi}_{\alpha'}^L)^t{\bf H}=E_{\alpha'} ({\vec
\Phi}_{\alpha'}^L)^t \label{left}
\end{equation}
By taking the transpose of Eq.{\ref{left}} one gets,
\begin{equation}
{\bf H}^t{\vec \Phi}_{\alpha'}^L=E_{\alpha'} {\vec
\Phi}_{\alpha'}^L \label{left2}
\end{equation}
The matrix $\bf H$ and its transpose ${\bf H}^t$ support the same
spectrum. We define the generalized inner product, termed
c-product \cite{Mol-Phys-79}, such that for non degenerate states
where $E_{\alpha'}\ne E_{\alpha}$,
\begin{equation}
({\vec \Phi}_{\alpha'} |{\vec \Phi}_\alpha)\equiv ({\vec
\Phi}_{\alpha'}^L)^t\cdot {\vec \Phi}_\alpha^R  = 0
\end{equation}
From Eq.\ref{left2} it is obvious that when $\bf H$ is Hermitian
matrix (i.e., ${\bf H}^t={\bf H}^*$) then ${\vec
\Phi}_{\alpha'}^L=({\vec \Phi}_{\alpha'}^R)^*$ and the c-product
$(...|...)$ is equal to the known scalar product $<...|... >$. If
for example $\bf H$ is a symmetric matrix (note any matrix can be
transformed to a symmetric form and therefore a complex symmetric
matrix is a general representation of a NH matrix) then,
\begin{equation}
{\vec \Phi}_{\alpha}^L={\vec \Phi}_{\alpha}^R
\end{equation}
Do the eigenvectors of a non-hermitian matrix $\bf H$ form a
complete set ? A complete set implies that the number of linearly
independent eigenvectors is equal to the dimension of the matrix.
It may happen that due to a coalescence of two eigenvectors
(coalescence of more than two eigenvectors is very unlikely
phenomena\cite{wilkinson}\cite{NM-SF-JCP}) the number of the
linearly independent eigenvectors is smaller than the dimension of
the matrix and the spectrum is incomplete. This coalescence of the
eigenvalues happens at $\alpha=\alpha_b$ (where "b" stands for a
branch point in the complex energy plane\cite{NM-SF-JCP})for
$\alpha \to \alpha_b$ the specific eigenvector denoted by
$\alpha=\alpha_b$  is "self-orthogonal" in the sense that, $({\vec
\Phi}_{\alpha_b}|{\vec \Phi}_{\alpha_b})= 0$. However, due to
round off numerical errors it is quite impossible to get the
incomplete spectrum and always in the numerical calculations the
eigenvectors are normalizable such that. $({\vec \Phi}_{\alpha}
|{\vec \Phi}_\alpha) = 1$ \cite{EN-NM}. As the number of the basis
functions, $\{\varphi_i\}$, which are used to construct the
Hamiltonian matrix in Eq.\ref{HAM})is increased, the function
$\Phi_\alpha^R(N)$ given by:
\begin{equation}
\Phi_\alpha^R(N)=\sum_{i=1}^N [{\vec \Phi}_\alpha^R]_i \varphi_i
\end{equation}
gives a better description of the $\alpha-th$ eigenfunction of the
Hamiltonian. It is out of the scope of this paper to  describe the
complex (i.e., NH) variational theorem that has been proved for
complex symmetric Hamiltonian matrices (see Ref.\cite{NM-review}
and references therein). However, on the basis of the complex
variational principle the exact eigenfunction, $\phi_\alpha^R$ can
be expressed as,
\begin{equation}
\phi_\alpha^R=\lim_{N\to\infty} \Phi_\alpha^R(N)
\end{equation}
and similarly,
\begin{equation}
\phi_\alpha^L=\lim_{N\to\infty} \Phi_\alpha^L(N)=\lim_{N\to\infty}
\sum_{i=1}^N [{\vec \Phi}_\alpha^L]_i \varphi_i^*
\end{equation}
where,
\begin{equation}
{\hat H} \phi_\alpha^R= E_\alpha  \phi_\alpha^R \label{phi_R}
\end{equation}
and
\begin{equation}
{\hat H}^\dag \phi_\alpha^{L*}= E_\alpha^* \phi_\alpha^{L*}
\label{phi_L}
\end{equation}
Here ${\hat H}^\dag$ is the hermitian conjugate of $\hat H$ such
that,
\begin{equation}
<\varphi_i|{\hat H}^\dag|\varphi_j>=[<\varphi_j|{\hat
H}|\varphi_i>]^*
\end{equation}
The basis functions are square integrable functions as described
above. Now, the generalized inner product (so called c-product)
for the left and right eigenfunction,
$\{\phi_{\alpha}^{L},\phi_\alpha^R\}$, will be defined by:
\begin{equation}
(\phi_{\alpha'}|\phi_\alpha)\equiv \int_{all-space}
\phi_{\alpha'}^{L}\phi_\alpha^R dv =\delta_{\alpha',\alpha}
\end{equation}
Note that the c-product, $(\phi_{\alpha}|\phi_\alpha) $, may be
complex and may have negative real or imaginary parts, therefore
the c-product is not a metric scalar product \cite{Mol-Phys-79}.
For a physical interpretation of the complex density probability
$\phi_{\alpha}^{L}\phi_\alpha^R$ see Ref.\cite{NM-Hadas}.
\subsection{Brief account of the  time asymmetry problem in NH-QM}
\label{asymmetry} Time asymmetry in physics is a concept closely
related to irreversibility which has been a subject of many
theoretical studies. The conventional Hermitian QM is
time-symmetric in the sense that it is described by an equation
symmetrical with respect to time and by time-symmetric boundary
conditions\cite{Bohm}. The process of the decay of a metastable
state is an irreversible one in the sense that it can only be
described from a certain time $t=0$ to time $t=\infty$. Authors
such as Bohm \cite{Bohm}, Prigogine \cite{Prigogine}, Sudarshan
et. al. \cite{Sudarshan} have constructed formalisms which
incorporate the irreversible nature of the resonance phenomena
into QM usually by relying on the introduction of rigged Hilbert
spaces. This leads to semi group evolution which distinguished
between "prepared" stated and "measured" states \cite{Bohm}.
However, it has been argued by Nicolaides \cite{Nicolaides} that
time asymmetry results from complex energy distribution without
the need to introduce rigged
Hilbert spaces.\\
Indeed, in the spirit of Ref.\cite{Nicolaides}, we will show that
in NH-QM, within the spectral representation of the NH
Hamiltonian, time asymmetry will pose a problem in the propagation
of WP's. The need to carry out WP calculations within the
framework of NH-QM is a crucial point in the study of systems
where the dynamics are not controlled by a single resonance state.
When one tries to propagate a WP in NH-QM the evolution of each of
the stationary eigenfunctions in Eq.\ref{phi_R} is governed by the
NH-TDSE and the stationary solutions of Eq.\ref{NH-TDSE} are given
by:
\begin{equation}
\psi_\alpha^R(t)=\exp(-\imath E_\alpha t/\hbar)\phi_\alpha^R
\quad. \label{psi_R}
\end{equation}
When one attempts to propagate the corresponding stationary
solutions of the transposed NH Hamiltonian
$\hat{H}^t=\hat{H}^{\dag *}$ (see Eq.\ref{phi_L}), at time $t$ the
left eigenfunctions will be given by:
\begin{equation}
\psi^L_\alpha(t)=\exp (+iE_\alpha t/ \hbar)\phi_\alpha^L \quad.
\label{psi_L}
\end{equation}
Since $E_\alpha= {\cal E}_\alpha -\imath \Gamma_\alpha / 2$ the
left eigenfunctions diverge exponentially as $t\rightarrow
\infty$. In Hermitian QM $\psi^L_\alpha=\psi^*_\alpha$ and
$E_\alpha$ is real, thus, $\psi^L_\alpha$ does not diverge. Since
the only difference from the evolution of $\psi^R_\alpha$ is in
the $+$ sign in the exponent it is similar in a way to propagating
the initial state to a negative time. Since in NH-QM the energy is
complex, the wavefunction cannot be propagated backward to $-
\infty$ but only forward to $+\infty$. This time asymmetry problem
dissolves when studying cases where only one metastable state
dominates since when calculating any observable quantity the time
dependent phases of the right and left eigenfunctions, $\exp(\pm
\imath E_\alpha t/\hbar)$, cancel out. When one studies the
evolution of a WP, the right function obeys the NH-TDSE in
Eq.\ref{NH-TDSE} and can be constructed as a linear combination of
the eigenfunctions of the NH Hamiltonian given in Eq.\ref{psi_R}.
The left function will be a linear combination of the
eigenfunctions given in Eq.\ref{psi_L}. Now, when calculating any
observable quantity, which does not commute with $\hat{H}$, there
will be cross terms between different eigenfunctions with time
dependent phase of the form, $\exp(\imath
(E_\alpha-E_{\alpha'})t/\hbar)$. These terms will diverge at long
times as $\exp((\Gamma_\alpha-\Gamma_{\alpha'})t/2)$ and prevent
the
calculation of observable quantities of a WP in NH-QM.\\
 This is, in short, the time asymmetry problem in NH-QM. Due to this problem the
calculation of observables in NH-QM is feasible only when one long
lived metastable state controls the dynamics of the system and
$\psi_\alpha^L(t)\psi_\alpha^R(t)=\phi_\alpha^L\phi_\alpha^R$ is
time independent. A definition of an observable for a general
superposition of resonance states is required. This subject will
be discussed in Sections \ref{F-product} and \ref{1Dmodel}.
\section{Wavepacket propagation in the F-Product formalism}
\label{F-product} As stated earlier the time asymmetry problem
prevents the propagation of WP's backward in time in NH-QM. The
existing definition of the left eigenfunctions of the NH
Hamiltonian is based on the c-product as defined in section
\ref{c-product} by Eq.\ref{psi_L}, which diverges exponentially in
time due to the complex part of the energy. The one-state
probability density
$\psi_{\alpha}^L\psi_{\alpha}^R=\phi_{\alpha}^L\phi_{\alpha}^R$ is
complex and time independent. However, as mentioned before the
imaginary part of the complex energy is associated with the decay
of a metastable state therefore one would expect that the
probability to find the particle in such a state should decay in
time with a rate of decay $\Gamma_\alpha$.\\
In order to impose this decay behavior of a metstable state we
apply the finite-range F-product (FP) formalism which was proposed
in Ref.\cite{NM-Lein} but was never tested by carrying out WP
propagation calculations within the framework of NH-QM. For the
sake of clarity let us first explain the idea behind the F-product
formalism. As we have seen in section \ref{asymmetry}, when the
left function is  {\it stationary } solution of Eq.~\ref{phi_L} it
is defined by Eq.\ref{psi_L}. Although $\psi_{\alpha}^L(t)$
diverges exponentially as time goes to infinity, the norm is
preserved and $(\psi_{\alpha}(t)|\psi_{\alpha}(t))=(\phi_{\alpha}
| \phi_{\alpha})=1$. This result is due to the fact that we
integrate over the {\it entire} space and the probability to find
the particle somewhere in space is unity. In order for the
probability density to decay with time we need to divide the
entire space into two parts. One part is the interaction region
which we associate with our system that in time breaks up into
sub-systems. The complimentary part is defined as a "surrounding"
or as the "environment" of our system. We require from an isolated
single metastable state to decay exponentially (a first order
'reaction')  n time as the sub-systems
escape from the interaction region to the "environment".\\
An acceptable way to separate the "system" from the "surrounding"
is by the well known Feshbach formalism \cite{Feshbach}, which is
often used for describing systems in nuclear physics. Here the
Hamiltonian is split by two operators $Q$ and $P$ which project
the system into subspaces of discrete and continuum states
respectively. The resulting Hamiltonian for the bound part of the
system is given by:
\begin{equation}
\label{FeshHam}
\hat{H}_{FF}=\hat{H}_{QQ}+V_{QP}\hat{G}_P^{+}V_{PQ} \quad,
\end{equation}
where, $\hat{H}_{QQ}$ is the Hamiltonian for the "system" and
$\hat{G}_P^{+}=\lim_{\epsilon\to
0^+}(E-\hat{H}_{PP}+i\epsilon)^{-1}$ is the Green operator for the
particle in the continuous or dissociative part of the spectrum
which describes the "surrounding" to which the particles decay.
The effective Hamiltonian as defined in Eq.~\ref{FeshHam} is a NH
operator. It is clear that here the effective Hamiltonian is NH
due to splitting of space into an interaction region where our
system is located and a "surrounding" which absorbs the emitted
particles (sub-systems). However, we wish to associate the decay
phenomena with particles which "escape" from the interaction
region to its "surrounding" even in cases where the NH formalism
takes the entire space into account. How to split the entire space
in the most general case into an interaction region and its
"surrounding" ? We will return to this question in the next
section. Here we argue that when the dynamics is controlled by a
single isolated metastable state within the NH-QM formalism then
we may use the finite box quantization approach where the
interaction region can be defined as a box as large as one wishes.
In such a case the integrals are calculated not over the entire
space, i.e., from $-\infty$ to $+\infty$, but from $-a/2$ to
$+a/2$ where the size of the box, $a$, is as large as one wishes,
hence the name - finite-range given to this new formalism. Since
we want to describe a decay phenomena we now define the evolution
of the left eigenfunctions as,
\begin{equation}
\psi_{\alpha}^L=\exp(\imath E_{\alpha}^* t/\hbar)\phi_{\alpha}^L=
\exp(\imath \varepsilon_{\alpha}t/\hbar)
\exp(-\Gamma_{\alpha}t/(2\hbar))\phi_{\alpha}^L \quad .
\end{equation}
Now the probability do detect the particle in a decaying state
will decrease exponentially in time according to:
\begin{equation}
(\psi_{\alpha}|\psi_{\alpha})=\exp(-\Gamma_{\alpha} t/\hbar)
(\phi_{\alpha}|\phi_{\alpha})=\exp(-\Gamma_{\alpha} t/\hbar)
\quad.
\end{equation}
This definition also prevents the divergence in time of the left
functions  and enables the propagation of a WP in time. Within the
FP formalism a WP will be now defined by a superposition of the
eigenfunctions of the NH Hamiltonian.
\begin{equation}
\label{FP-psi} \Psi^R_{FP}=\sum_\alpha C_\alpha^R \exp (-\imath
E_\alpha t/\hbar)\phi_\alpha^R \qquad ; \qquad
\Psi^L_{FP}=\sum_\alpha C_\alpha^L \exp (\imath E_\alpha^*
t/\hbar)\phi_\alpha^L \quad ,
\end{equation}
where
\begin{equation}
C_\alpha^R=\left(\phi_\alpha|\Psi_{FP}(t=0)\right) \qquad; \qquad
C_\alpha^L=\left(\Psi_{FP}(t=0)|\phi_\alpha\right) \quad.
\end{equation}
When the Hamiltonian is NH due to the application of one of the
complex scaling (CS) similarity transformations, then the initial
state $\Psi_{FP} (t=0)$ is given by ${\hat S}\Psi(t=0)$ where
$\hat S$ is one of the CS transformations\cite{NM-Hirschfelder}.
On the other hand when CAP is added to the Hamiltonian and the
initial state is localized in the interaction region it remains
unscaled. An important point in the new definition of the inner
product is that now the right-WP satisfies the NH-TDSE as defined
in Eq.\ref{NH-TDSE}, whereas the left-WP does {\it not} satisfy a
corresponding wave equation. This is a very important point in our
search for solution of the time asymmetric problem in NH-QM. The
FP definition of the inner product generalizes only the time
dependence of the c-product definition. Therefore, when the
functions are time independent the F-product reduces to the
c-product. Based on the Eq. \ref{FP-psi} the decay of a WP will
now be given by:
\begin{equation}
\left(\Psi_{FP}|\Psi_{FP} \right)=\sum_{\alpha} C_\alpha^L
C_\alpha^R \exp\left(-\Gamma_\alpha t/\hbar\right).
\end{equation}
Note that although there are no cross terms between different
$\alpha$'s there are still interference effects since
$C_\alpha^LC_\alpha^R$ have complex values. Nevertheless the
absolute value of this expression decays continuously as time
passes. The goal of this paper is to check the FP formalism for
the first time for a simple test model problem and compare the
results with those obtained from conventional QM calculations.
Such a comparison is made in section \ref{1Dmodel}.
\section{Application to a simple one dimensional problem}
\label{1Dmodel} 
\begin{figure}
\begin{center}
 \epsfxsize=10 cm \epsffile{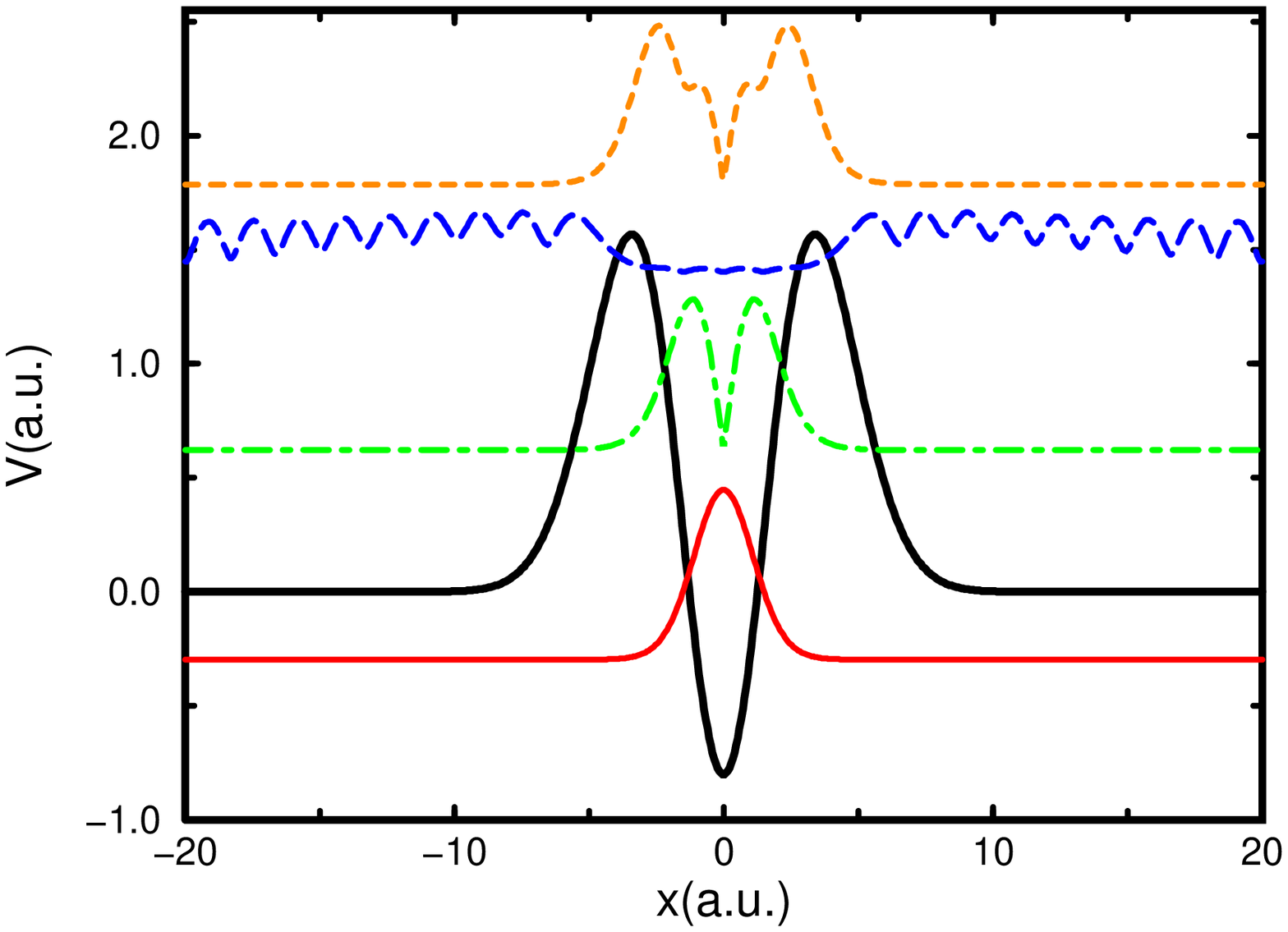}
\caption{\label{JOLA-CS}(Color Online) Eigenfunctions of the model
potential in Eq.\ref{jolanta-CS}. The solid line is for the bound
state; the dot-dashed line is for an isolated resonance; The long
dashed line is for a continuum state just under the top of the
barrier; and the dashed line is for a broad resonance over the
barrier.}
\end{center}
\end{figure}

\begin{figure}
\begin{center}
\epsfxsize=10 cm \epsffile{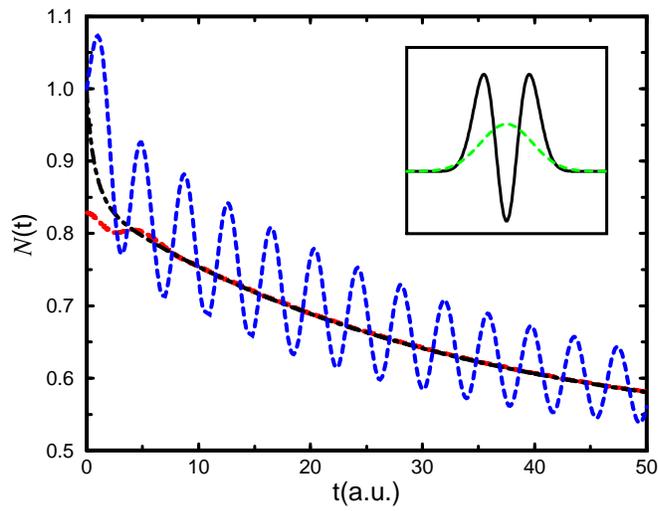} \caption{\label{norm1}
(Color Online) The norm of  wide Gaussian WP based on the
F-product formalism as given by Eq.\ref{normFP} (dot-dashed
line)as a function of time in comparison with the part of the
Hermitian WP localized inside the interaction region (long-dashed
line)as given in Eq.\ref{normHQM}. The dashed line shows the norm
in the formalism portrayed in Eq.\ref{norm_np}. In the inset the
initial gaussian WP with width of $\sigma=3.87a.u.$ and
$k_0=0a.u.$ .}
\end{center}
\end{figure}

\begin{figure}
\begin{center}
 \epsfxsize=10 cm \epsffile{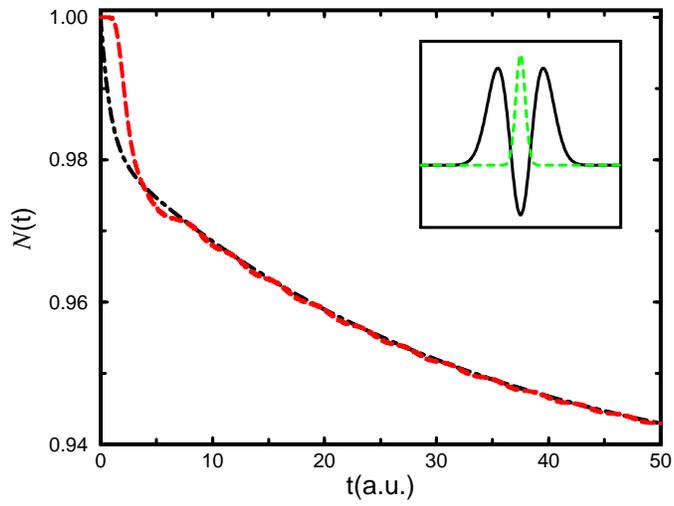} \caption{\label{norm2} (Color
Online) The same as Fig\ref{norm1}, but with an initial state of a
narrow gaussian, $\sigma=0.71a.u.$, fully localized in the
interaction region around $x=0$ (shown in the inset). The
dot-dashed line is for Eq.\ref{normFP} and the long dashed line
for Eq.\ref{normHQM}.}
\end{center}
\end{figure}

\begin{figure}
\begin{center}
\epsfxsize=10 cm \epsffile{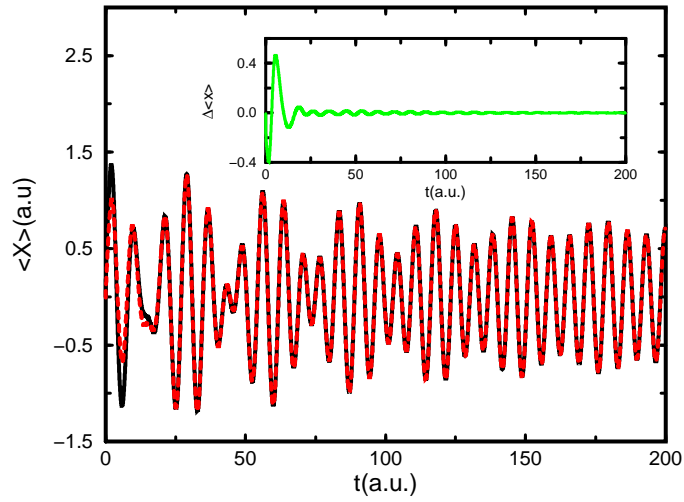} \caption{\label{aveX}(Color
Online) The average position of a gaussian WP with initial
momentum of $k=1a.u.$ and $\sigma=3.87a.u.$ as a function of time.
The solid line is for the solution in the F-product formalism as
given in Eq.\ref{FPobs}while the dashed line is for the Hermitian
part of the WP localized inside the interaction region as depicted
in Eq.\ref{IRobs}. The inset shows the difference between the two
as a function of time.}
\end{center}
\end{figure}

\begin{figure}
\begin{center}
 \epsfxsize=10 cm \epsffile{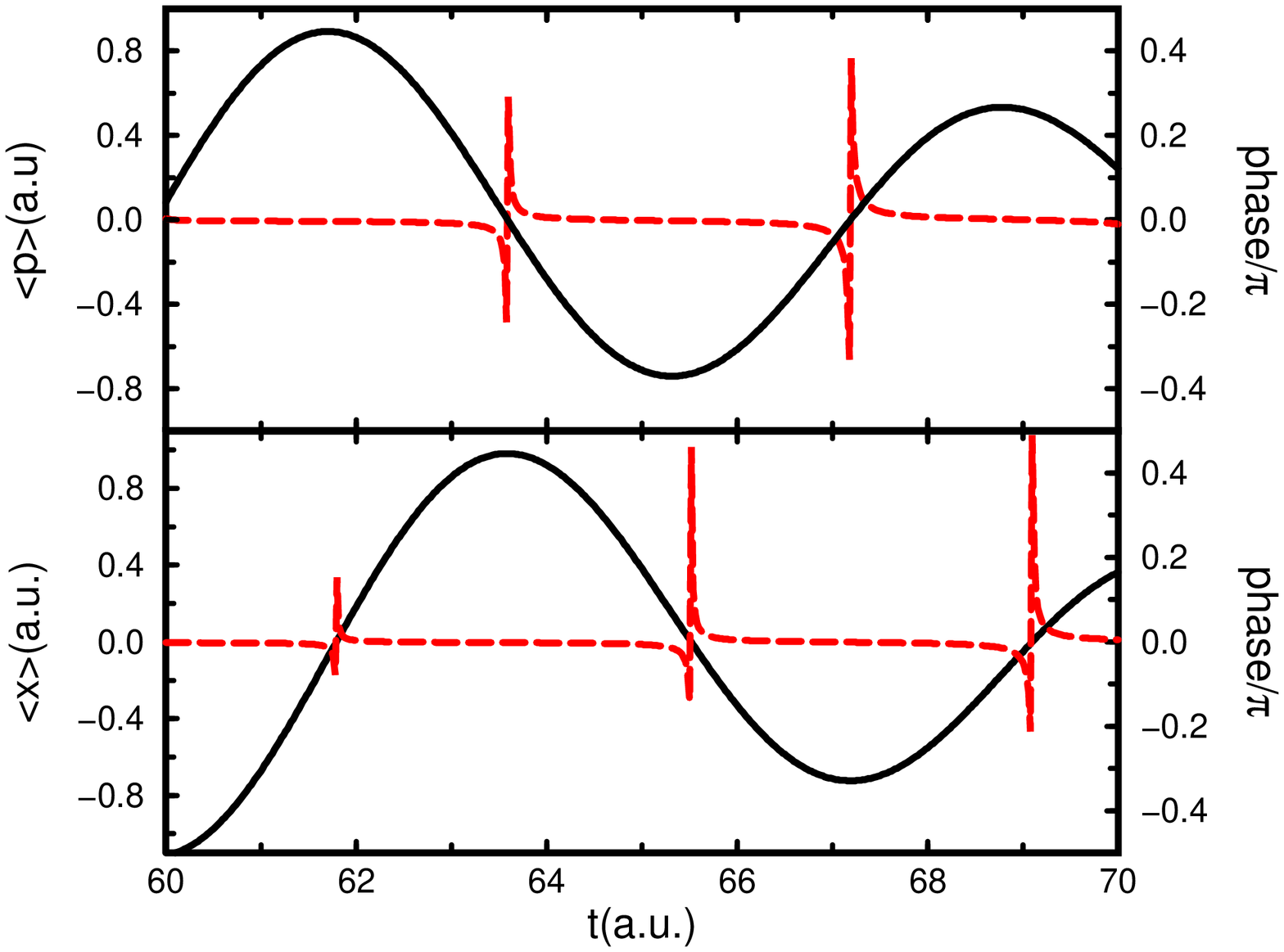}
\caption{\label{xp-phase} (Color Online) The average position
(bottom) and momentum (top) of a gaussian WP with initial momentum
$k=1$ in the F-Product formalism and their corresponding phases
based on Eq.\ref{FPobs}. The solid line is for the real part of
the average position/momentum while their corresponding  phases
are given by the long-dashed lines. The interval of time on which
these results appear was chosen arbitrarily in order to have clear
presentation, but the overall behavior is similar in all range of
time. Due to the abruptness of the change the shifts in pahsesd
are not exactly of $\pi$ strictly for numerical reasons.}
\end{center}
\end{figure}

We would like to implement the new formalism on a
simple one dimensional time independent Hamiltonian, which will
serve as a test model:
\begin{equation}
\hat{H}(x)=\frac{\hat{p}^2}{2m}+(\frac{x^2}{2}-0.8)\exp(-0.1x^2)\quad,
\label{jolanta}
\end{equation}
This model Hamiltonian has been often used as a test case for new
theories and computational methods(see for example
\cite{jola-refs}). The potential in Eq.\ref{jolanta} consists of a
potential between two barriers (See Fig.\ref{JOLA-CS}). In the
example presented below the Hamiltonian becomes NH upon complex
scaling of the coordinate by a complex phase, $\theta$, such that
$x \to x\exp(\imath \theta)$ and is given by:
\begin{equation}
\hat{H}_\theta(x)=e^{-2\imath\theta}\frac{\hat{p}^2}{2m}+
(\frac{e^{2\imath\theta}x^2}{2}-0.8)\exp(-0.1e^{2\imath\theta}x^2)\quad,
\label{jolanta-CS}
\end{equation}
The model potential shown in Fig.\ref{JOLA-CS} supports a bound
state and two metastable resonances below the top of the barrier
as well as several other resonances over the barrier. The
resonances are localized states that decay in time and cannot be
described by a single eigenfunction of the Hermitian Hamiltonian
in Eq.\ref{jolanta}, however in NH-QM they are represented by a
single square integrable eigenfunction of the NH Hamiltonian in
Eq.\ref{jolanta-CS} as can be seen in Fig.\ref{JOLA-CS}. Our
objective is to show that the FP formalism will yield similar
results to those obtained by conventional Hermitian QM, without
the need to divide our space to "system" and "surrounding". For
this purpose lets define now the interaction region as the area
between the top off the two barriers in Fig.\ref{JOLA-CS} and
return to this point later on.\\
In order to observe the decay of a WP with time we will place a
gaussian WP in the center of the potential in Eq.\ref{jolanta} of
the form:
\begin{equation}
\label{gaussian} \Psi(0)=\frac{1}{(\pi\sigma^2)^{1/4}}
\exp\left(-\frac{x^2}{2\sigma^2}+\imath k_0 x\right) \quad,
\end{equation}
where $\sigma$ is the width of the WP and $k_0$ its initial
momentum. Upon CS the right initial state WP becomes:
\begin{equation}
\Psi^R(0)=\frac{e^{\imath\theta}}{(\pi\sigma^2)^{1/4}}
\exp\left(-\frac{e^{2\imath\theta}x^2}{2\sigma^2}+\imath
e^{\imath\theta}k_0 x\right) \quad,
\end{equation}
while the left initial WP is given by:
\begin{equation}
\Psi^L(0)=\frac{e^{\imath\theta}}{(\pi\sigma^2)^{1/4}}
\exp\left(-\frac{e^{2\imath\theta}x^2}{2\sigma^2}-\imath
e^{\imath\theta}k_0 x\right)\quad.
\end{equation}
Note that at time $t=0$ the definitions of the c-product and the
F-product are identical and $\left(\Psi^L(0)|\Psi^R(0)\right)=1$.
The norm of the WP, ${\cal N}_{FP}(t)$, within the FP formalism,
as defined in Eq.\ref{FP-psi}, and is given by:
\begin{equation}
\label{normFP}
 {\cal N}_{FP}(t)=\int_{\infty}^{\infty}\Psi^L_{FP}(x,t)\Psi^R_{FP}(x,t)dx=\sum_\alpha
C_\alpha^L C_\alpha^R \exp(-\Gamma_\alpha t/ \hbar) \quad .
\end{equation}
Assuming the decay of the WP is a first order process the
effective decay rate will be given by:
\begin{equation}
k_{FP}(t)=-\frac{d}{dt}\ln {\cal N}_{FP}(t) \quad.
\end{equation}
When the WP populates several resonance states the effective decay
rate will be time dependent, but when only one resonance state is
populated the anticipated constant effective decay rate will be
obtained, $k_{FP}=\Gamma_{res}/\hbar$. In contrast, the decay will
not be observed in Hermitian QM where the norm is conserved.
Therefore we choose a new quantity for comparison with the norm in
Eq.\ref{normFP}, which we will label ${\cal N}_{QM}$ and is
defined by the part of the WP which is localized inside the
interaction region.
\begin{equation}
 {\cal N}_{QM}(t)=\int_{-a/2}^{a/2}\Psi^{*}_{QM}(x,t)\Psi_{QM}(x,t)dx \quad,
\label{normHQM}
\end{equation}
where $\Psi_{QM}(x,t)$ is the result obtained from conventional
propagation calculations, when by conventional we mean the
solution of the Hermitian TDSE. This enables us to define an
effective decay rate even in Hermitian QM based on
Eq.\ref{normHQM}.
\begin{equation}
k_{QM}(t)=-\frac{d}{dt}\ln {\cal N}_{QM}(t) \quad. \label{kQM}
\end{equation}
The question we address ourselves with now is how to define the
interaction region ? or similarly what is $a$ in Eq.\ref{normHQM}
? Is the initial assumption that we can define the interaction
region as the area between the top of the two barriers valid?
Here, we are not using the Feshbach formalism and therefore we
look for a simple universal definition. We propose here to define
the interaction region (the parameter "a" in our 1D case or the
vector $\vec a$ in multidimensional case) based on the NH
calculations. More precisely we wish to define the interaction
region as the region where the resonances are localized. The
continuum states in the same energy range have a very small
(almost vanishing ) amplitude in this region as can be seen in
Fig.\ref{JOLA-CS}. This definition is obviously not an exact one
and its important to note that using the FP formalism we avoid the
need to define the interaction region and thus $a$ is only
relevant when trying to relate the results to conventional QM. The
results which will be presented below strongly support
our conjecture. \\
Returning to the problem of the decay of the norm of a WP, we
first place a gaussian WP as shown in Eq.\ref{gaussian} with
$\sigma=3.87a.u.$ and $k_0=0$. It is evident on Fig.\ref{norm1}
that on the long time scale there is a remarkable correspondence
between the results of Eq.\ref{normFP} and Eq.\ref{normHQM}. On
short times there is deviation which results from the fact that
the decay of the resonances starts at time $t=0$ whereas for an
Hermitian WP it will take time to reach the boundary of the
interaction region. The idea is simple. Using the NH F-product
formalism the resonance states decay at any time including the
extremely short time regime. In conventional propagation
calculations the initial WP oscillates in between the barriers and
a significant part of the tunnelling takes place when the
oscillating WP hits the inner classical turning points of the
potential barriers. Therefore the deviation between the results
obtained from NH-QM and conventional QM calculations is during the
time it takes for the initial WP to reach the inner classical
turning points. To further illustrate this if we study a narrow WP
with $\sigma=0.71a.u.$, which is fully localized between the
barriers inside the potential in Fig. \ref{JOLA-CS} it is obvious
from the above argument that in Hermitian QM it will take the WP
some time to escape out of the barriers while in NH-QM decay
starts immediately at $t=0$ (see Fig\ref{norm2}). When one studies
the behavior of different WP's, the best choice for $a$ in
Eq.\ref{normHQM} varies and depends on the WP but it is always in
the vicinity of the top of the barriers in Fig.\ref{JOLA-CS}. This
can be understood again by realizing that the continuum
wavefunctions under the top of the barrier are localized outside
of the potential while the resonance states are localized inside
the potential (even over the barrier),
therefore, in NH-QM we have a clear distinction between the "system" and its "surrounding". \\
In  previous nuclear physics studies of the decay of narrow
isolated resonances which are associated with the NH effective
Hamiltonian given in Eq.\ref{FeshHam} the decay law has been
obtained by the conventional scalar product \cite{Rotter}. Namely,
following this approach the effective decay rate is given by:
\begin{equation}
k_{eff}(t)=-\frac{d}{dt}ln \left< \Psi^R(t) | \Psi^R(t) \right>
\quad,
\end{equation}
where $\Psi^R(t)$ is the solution of Eq.\ref{NH-TDSE} which can be
expanded in the basis set of the eigenfunctions of Eq.\ref{phi_R},
$\{\phi^R_\alpha\}$ (with corresponding eigenvalues $E_\alpha$),
such that $\Psi^R(t)=\sum_\alpha C_\alpha^R \phi^R_\alpha$. Using
this approach one gets that,
\begin{equation}
{\cal N}_{NP}=\left< \Psi^R(t) | \Psi^R(t) \right>
=\sum_{\alpha,\alpha'} (C_{\alpha'}^R)^{*}C_{\alpha}^R \exp
[-\imath(E_\alpha-E_{\alpha'})t/\hbar]\left< \phi^R_{\alpha'} |
\phi^R_\alpha \right> \quad,\label{norm_np}
\end{equation}
where the notation $NP$ stands for the formalism used in nuclear
physics in Ref.\cite{Rotter}. When the resonances are not isolated
there is a deviation from the effective decay rate due to
interference between the different states. When trying to apply
this approach to our simple test model, one fails to get converged
results and moreover the behavior of ${\cal N}_{NP}$ based on
Eq.\ref{norm_np}, as can be seen in Fig.\ref {norm1}, is quite
irregular as it oscillates instead of constantly decaying. This
shows that the formalism depicted in Eq.\ref{norm_np} can only be
used in specific cases whereas the FP formalism as
portrayed in Eq. \ref{normFP} should be used in general. \\
Let us now test the FP formalism for calculating time dependent
expectations values for NH Hamiltonians. If one wishes to measure
an observable quantity $\bar{A}$ which will be defined in NH-QM,
within the FP formalism, by:
\begin{equation}
\label{FPobs} \bar{A}(t)=\frac{\left(\Psi_{FP}(t)\left|
\hat{\tilde{A}}\right|\Psi_{FP}(t)\right)}{\left(\Psi_{FP}(t)|\Psi_{FP}(t)\right)}=
\frac{\int_{-\infty}^{\infty}
\Psi^{L}_{FP}(x,t)\hat{\tilde{A}}\Psi_{FP}^R(x,t)dx}
{\int_{-\infty}^{\infty}\Psi_{FP}^{L}(x,t)\Psi_{FP}^R(x,t)dx}
\quad,
\end{equation}
where in principle $\bar A$ can get complex values. In such a case
the phase of  $\bar A$ should be a measurable quantity (and
therefore should be $\theta$ independent although  both
$\Psi^R_{FP}(t)$ and $\Psi^L_{FP}(t)$ vary with $\theta$ (the
rotational angle associated with the CS transformation). In our
case (i.e., the Hamiltonian is NH due to the CS similarity
transformation.) $\hat{\tilde{A}}=\hat{S}\hat{A}\hat{S}^{-1}$ is
the scaled operator for the desired observable. As we have seen in
Fig's \ref{norm1},\ref{norm2} the NH formalism describes the part
of the WP which remains in the interaction region. Thus, we expect
to find correspondence to a quantity similar to that defined in
Eq.\ref{normHQM}. We will label $<A>$ as an observable quantity in
the interaction region calculated by the conventional (Hermitian)
quantum-mechanics approach.
\begin{equation}
\label{IRobs}
 <A(t)>=\frac{\int_{-a/2}^{a/2}
\Psi^{*}(x,t)\hat{A}\Psi(x,t)dx}{\int_{-a/2}^{a/2}\Psi^{*}(x,t)\Psi(x,t)dx}
\quad ,
\end{equation}
where $\Psi(x,t)$ is the solution obtained by solving the
conventional TDSE.  Returning to our one-dimensional test model,
the results for the observable $\hat{A}=x$ of the average position
for a WP placed at $x=0$ with width, $\sigma=3.87a.u.$, and
initial momentum $k=1a.u.$ based on Eq.'s \ref{FPobs},\ref{IRobs}
are given in Fig.\ref{aveX}. Once again there is very good
correspondence on the long time scale while on the short time
scale there are deviations. This suggests that the NH formalism
which describes the resonance states applies on times longer than
some initial rearrangement time in which the resonance states are
populated.\\
Since the scaling parameter is not associated with any physical
quantity, time dependent observables should get real values in
spite of the analytical continuation of the Hamiltonian. Indeed,
the results presented in Fig.\ref{xp-phase} show very clearly that
at any given time the mean position,
$\bar{x}(t)$,($\hat{A}=\hat{x}$ in Eq.\ref{FPobs}), and mean
momentum $\bar{p}(t)$,($\hat{A}=\hat{p}_x$ in Eq.\ref{FPobs}) are
real quantities even though they have been obtained by NH
calculation based on Eq.\ref{FPobs}. The phase of both the average
position and momentum is zero with abrupt shifts in $\pi$ every
time an observable vanishes (and thus can't be measured at that
given time). The jumps in $\pi$ of the phase of $\bar A$
($\bar{x}(t)$ or $\bar{p}(t)$) result from the fact that the
observable changes its sign from positive to negative or
vise-versa and therefore would appear also in standard  QM
calculations.
\section{Conclusions}
We have shown that using the recently introduced FP formalism it
is possible to overcome the time-asymmetry problem in NH-QM and to
propagate a WP which populates many resonance states. This
formalism describes the dynamics without the need to separate the
entire space into "system" and "surrounding" which is necessary
when using Hermitian QM. The NH description of the system is valid
on the timescales which are long enough to regard only the
localized part of the WP after the scattered part of the WP has
left the interaction region. It has also been shown that
observables such as position and momentum in the F-product
formalism obtain real values despite the NH nature of the
Hamiltonian. This reasserts the validity of this formalism which
makes it applicable to systems where particles are temporarily
trapped in the interaction region. The application of these novel
approach to many electron systems is under current study.
\newpage

\bigskip
\newpage

\begin{thebibliography}{99}

\bibitem{Junker-review}
B. R. Junker, Adv. At. Mol. Phys., {\bf 18}, 107 (1982).

\bibitem{Renihardt-review}
W. P. Reinhardt, Annu. Rev. Phys. Chem.,{\bf 33}, 223 (1982).

\bibitem{Ho-review}
Y. K. Ho, Phys. Rep. C, {\bf 99}, 1 (1983).

\bibitem{NM-review} N. Moiseyev, Phys. Rep. {\bf 302}, 211 (1998).

\bibitem{Caps} J. G. Muga, J. P. Palao, B. Navarro, and I. L.
Equsquiza, Phys. Report {\bf 395}, 357 (2004).

\bibitem{lenz-robin-review}
R. Santra and L. S. Cederbaum, Phys. Rep., {\bf 368}, 1 (2002).

\bibitem{zeldovich}
A. M. Perelomov and Ia. B. Ze\'{l}dovich, { \it Quantum Mechanics:
Selected Topics}, (World Scientific Publishing, Singapore, 1998).

\bibitem {Letropet}
E. Br\"{a}ndas and N. Elander (Editors), The Letropet Symposium
View on Generalized Inner Product, { \it Lecture Notes in Physics
vol. 325} (Springer, Berlin, 1989).

\bibitem{Mol-Phys-79}
N. Moiseyev, P. R. Certain, and F. Weinhold, Mol. Phys., {\bf 36},
1617 (1978).

\bibitem{NM-FW-PRL}
N. Moiseyev and F. Weinhold, Phys. Rev. Lett., {\bf 78}, 2100
(1997).

\bibitem{kosloff}
R. Kosloff and D. Kosloff, J. Comp. Phys., {\bf 63}, 363 (1986);
G. Ashkenazi, R. Kosloff, S. Ruhman, and H. Tal-Ezer, J. Chem.
Phys, {\bf 103}, 10005 (1995).

\bibitem{Leforestier-Wyatt}
C. Leforestier and R. E. Wyatt, Chem Phys. Lett.,{\bf 78}, 2334
(1983).

\bibitem{Balint-Kurti}
A, Vibok and G. G. Balint-Kurti, J. Chem. Phys., {\bf 96}, 7615
(1992);G. G. Balint-Kurti, Lecture Notes in Chemistry, {\bf 75},
74 (2000).

\bibitem{Dieter-Mayer}
E. Pahl, H. D. Meyer, L. S. Cederbaum, and F. Tarantelli, J. elec,
Spec. and Related Phenomena,{\bf 93}, 17 (1998).

\bibitem{NM-SS-LSC}
S. Scheit, L. S. Cederbaum, and H. D. Meyer, J. Chem. Phys., {\bf
118}, 2092 (2003); N. Moiseyev, S. Scheit, and L. S. Cederbaum,J.
Chem. Phys., {\bf 121}, 722 (2004).

\bibitem{Bohm}
A. Bohm, Phys. Rev. A, {\bf 60}, 861 (1999); A. Bohm and N.L.
Harshman, Lecture Notes in Physics, {\bf 504}, 181 (1998); A.
Bohm, Adv. Chem. Phys., {\bf 122}, 301(2002); A. Bohm et. al.,
Fort. der Phys., {\bf 51}, 551(2003).

\bibitem {Prigogine}
I. Prigogine, J. Int. Quantum Chem., {\bf 53}, 105 (1995).

\bibitem {Sudarshan}
E. C. G. Sudarshan, C. B. Chiu, and G. Bhamathi, Adv. Chem. Phys.,
{\bf 99}, 121 (1197).

\bibitem{Nicolaides}
C. A. Nicolaides, Lecture Notes in Physics, {\bf 622}, 357 (2003);
C. A. Nicolaides and D. R. Beck, Int. J. Quantum Chem., {\bf
14},457 (1978); C. A. Nicolaides, Int. J. Quantum Chem., {\bf 89},
94 (2002) and references therein.

\bibitem{NM-Lein} N. Moiseyev and M. Lein, J. Phys.
Chem., {\bf 107}, 7181 (2003).

\bibitem{wilkinson} J. H. Wilkinson, {\it The Algebraic Eigenvalue
Problem}, (Oxford Clarendon Press, 1965).

\bibitem{NM-SF-JCP}
N. Moiseyev and S. Friedland, J. Chem Phys., {\bf 74}, 4739
(1981).

\bibitem{EN-NM}
E. Narevicius and N. Moiseyev, Prog. Theo. Chem. Phys.,{\bf 12},
311 (2003).

\bibitem{NM-Hadas}
H. Barkay and N. Moiseyev,Phys. Rev. A, {\bf 64}, 044702 (2001).

\bibitem{Feshbach}
H. Feshbach, Ann. Phys.,{\bf 5},357 (1958); H. Feshbach, {\it
Theoretical Nuclear Physics}(Wiley, New York, 1992).

\bibitem{NM-Hirschfelder}
N. Moiseyev, J. O. Hirschfelder, J. Chem. Phys., {\bf 88}, 1063
(1998).

\bibitem{jola-refs}
N. Moiseyev, P.R. Certain, and F. Weinhold, Mol. Phys., {\bf 36},
1613 (1978); N. Lipkin, N. Moiseyev, and E. Br\"{a}ndas, Phys.
Rev. A, {\bf 40},549 (1989);N. Moiseyev, Mol. Phys.,{\bf 47},585
(1982); H. J. Korsch, H. Laurent, R. M\"{o}hlenkamp, Mol. Phys.,
{\bf 43}, 1441 (1981); M. Rittby, N. Elander, E. Br\"{a}ndas,
Phys. Rev. A, {\bf 24}, 1636 (1981).

\bibitem{Rotter}
E. Persson, T. Gorin, and I. Rotter, Phys. Rev. E, {\bf 54}, 3339
(1996).


\end{thebibliography}
\end{document}